\documentclass[12pt]{article}
\usepackage{amsmath,amssymb}
\usepackage{amsmath,amssymb,epsf}
\newcommand{\be}{\begin{eqnarray}}
\newcommand{\ee}{\end{eqnarray}}
\newcommand{\ba}{\begin{array}}
\newcommand{\ea}{\end{array}}

\def\bbox{{\,\lower0.9pt\vbox{\hrule \hbox{\vrule height 0.2 cm
\hskip 0.2 cm \vrule height 0.2 cm}\hrule}\,}}
\newcommand{\dsl}{\pa \kern-0.5em /}

\begin{document}

%%%%%%%%%%%%%%%% title page %%%%%%%%%%%%%%%%%%%%%%%%%%%%%%%%%%%%

\begin{titlepage}

\vfill

\begin{center}
\baselineskip=16pt {\Large Non-integer flux --- why it does not work.}
\vskip 0.3cm
{\large {\sl }}
\vskip 10.mm
{\bf 
A.~V. Smilga}
 \\

\vspace{6pt}
SUBATECH, Universit\'e de Nantes, \\
4 rue Alfred Kastler, BP 20722, Nantes 44307, France
\footnote {On leave of absence from ITEP, Moscow, Russia.}

\end{center}
\vspace{2cm}

\par
\begin{center}
{\bf ABSTRACT}
\end{center}
\begin{quote}

We consider the Dirac operator on $S^2$ without one point in the case of non-integer 
magnetic flux. We show that the spectral problem for $H = /\!\!\!\!{\cal D}^2$
can be well defined, if including in the Hilbert space ${\cal H}$ only nonsingular 
on $S^2$ wave functions. However, 
this Hilbert space is not invariant under the action of $ /\!\!\!\!{\cal D}$ --- 
for certain $\psi \in {\cal H}$, \ $/\!\!\!\!{\cal D} \psi$ does not belong to ${\cal H}$ anymore.
This breaks explicitly the supersymmetry of the spectrum. In the integer flux case, supersymmetry
can be restored if extending the Hilbert space to include locally regular sections of the
corresponding fiber bundle. For non-integer fluxes, such an extension is not possible.
\end{quote}
\end{titlepage}
%%%%%%%%%%%%%%%%%%%%%%%%%%%%%%%%%%%%%%%%

\setcounter{equation}{0}
\section{Introduction}
  The famous Atiyah-Singer theorem \cite{AS} relates the index of the Dirac operator $/\!\!\!\!{\cal D}$ 
(the difference $n^{(0)}_L - n^{(0)}_R$, $n^{(0)}_{L,R}$ being the number of left-handed (right-handed) 
zero modes of \ $/\!\!\!\!{\cal D}$) on a manifold ${\cal M}$ to certain topological invariants. The most important of  these
 invariants are the   Chern classes of the 
corresponding fiber bundles,
 \be
\label{Chern} 
{\rm Ch}(F)\ \propto {\rm Tr} \int_{\cal M} \, F\wedge \ldots \wedge F \ .
 \ee
($F = (1/2)F_{\mu\nu} dx^\mu \wedge dx^\nu$). In the simplest nontrivial case of $U(1)$ bundle on $S^2$, 
the index coincides with the magnetic flux,
 \be
\label{IndS2}
 I \ =\ \frac 1{2\pi} \int F \ .
 \ee

When a mathematician talks about a fiber bundle on $S^2$, he has in mind a particular construction \cite{EGH}. 
The sphere is covered by two overlapping 
maps with the coordinates  $\{x_\mu\}$ and $\{x'_\mu\}$. On each map, 
the connections $A = A_\mu dx^\mu$ and     $A' = A'_\mu dx'^\mu$ are defined.
One requires then that, in the region where the maps overlap, the connections $A$ and $A'$ are related by a gauge transformation,
 \be
\label{AAprim}
A'_\mu = \Omega^{-1}(x) [A_\mu +i \partial_\mu] \Omega(x)\ ,
 \ee
 where $\Omega(x) \in U(1)$ is a well-defined function in the overlap region. 
With this definition, the
magnetic flux (\ref{IndS2}) is always integer. 

A convenient explicit realisation of this construction involves two sets of stereographic coordinates  $\{x, y\}$ and $\{x', y'\}$.
One of the maps covers the whole $S^2$ but the north pole and another --- $S^2$ but the south pole. The relationship 
(\ref{AAprim})  can be established on the equator with topology $S^1$. 
The magnetic flux coincides then with the winding number of the
map $U(1) \to S^1$. 

One can still pose a naive question. Suppose we consider the Dirac equation on only one such stereographic map, 
disregarding the singularity at the pole. The magnetic flux can then acquire an arbitrary value. The question is whether the spectral
problem with a fractional flux 
defined on $S^2 \backslash $\{{\sl one point}\} is benign and, if not, what in particular goes wrong there.

The answer is the following. 
\begin{itemize}
\item
To begin with, when the north pole of the sphere (corresponding to $x = y = \infty$) is disregarded,
 it seems  natural to include into consideration the wave functions that are singular at this point and only require for the functions
to be square integrable. It turns out, however, that, for non-integer fluxes,
 the Hamiltonian $H = /\!\!\!\!{\cal D}^2$ {\it is} not Hermitian in 
such extended Hilbert space.

\item We can impose the requirement for the function to be regular at all points of $S^2$ including the north pole. In this case, 
the Hamiltonian is Hermitian. However, the Dirac operator itself is not well defined in the framework of such  reduced Hilbert space 
${\cal H}$: {\it some} nonsingular on $S^2$ functions become singular when acted upon by  $/\!\!\!\!{\cal D}$. 
A mathematician
would say that this Hilbert space does not constitute a {\it domain} of \ $/\!\!\!\!{\cal D}$. A physicist would notice that
certain regular on $S^2$ functions do not have regular superpartners (obtained by the action
 of the supercharges $Q, \bar Q = /\!\!\!\!{\cal D} (1 \pm \sigma^3)/2$). And that means supersymmetry breaking. 
 \item 
The remark above (that supersymmetry is broken if including into consideration only regular functions) refers both to
integer and fractional fluxes. However, if the flux is integer, supersymmetry of the Hamiltonian
$H = /\!\!\!\!{\cal D}^2$ can be restored if including into the Hilbert
space square integrable functions which behave as  $ \psi \sim e^{i\phi}$ in the vicinity of the north pole (such that
$|\psi| = C \neq 0$). They are nothing but the sections of the fiber bundle described above and 
restricted on the map not covering
the north pole.
 For noninteger fluxes, this is not possible, which means that supersymmetry is genuinely broken.  
 \end{itemize}

We guess that the same happens in all other cases when one tries to consider the Dirac spectral problem for ``crippled'' bundles with 
non-integer Chern classes. 

Note that we discuss in this paper only compact manifolds. If the manifold is not compact (for example, $R^2$), the spectrum of 
$/\!\!\!\!{\cal D}^2$ is continuous  \cite{Casher}. And continuous spectrum is not quite the {\it spectrum} in a rigid mathematical sense. 
To pose the questions
about hermiticity, etc., one has to regularize the problem in the infrared rendering the motion finite and the spectrum discrete. 
One of the ways to perform such a regularisation is to consider, instead of a non-compact manifold, a manifold with a boundary. 
Then supersymmetry
can be preserved by imposing rather complicated {\it nonlocal} boundary conditions due to Atiyah, Pathody and Singer 
\cite{APS}. We will not discuss
this further.  

\section{Monopole harmonics}
We choose the complex stereographic coordinates $ w,\bar w$. The metric is
 \be
\label{metrS2}
ds^2 \ =\ \frac {2d w d \bar w}{(1 + \bar w w)^2}\ .
  \ee
It is singular at $w=\infty$, but this singularity is integrable. In particular, the volume of such sphere is
 \be
\label{VS2}
V = \ \int \frac {d w d\bar w}{(1+\bar w w)^2} \ =\ 2\pi \, ,
 \ee
corresponding to the radius $R = \sqrt{2}/2$. Consider the action \cite{IvSm}
 \be
\label{actionIS}
S = \int dt d^2\theta \left[ - \frac {DW \bar D \bar W}{4(1 + \bar W W)^2} + G(\bar W, W) \right]\ ,
 \ee
where
$$ {\cal D} = \frac {\partial}{\partial \theta } - i \bar \theta \frac {\partial}{\partial t }\, ,
 \ \ \ \ \ \ \ \ \  \bar { \cal D} = -\frac {\partial}{\partial \bar\theta } + i  \theta \frac {\partial}{\partial t }$$
are supersymmetric covariant derivatives, 
 \be
\label{WWbar}
W \ =\ w + \sqrt{2} \theta \psi - i\theta \bar \theta \dot{w}, \ \ \ \ 
\bar W \ = \ \bar w - \sqrt{2} \bar\theta \bar\psi + i\theta \bar \theta \dot{\bar w}
  \ee
are chiral $(0+1)-$dimensional superfields, $\bar D W = D \bar W = 0$, and $G(\bar W, W)$ is an arbitrary
real 
function. 
After deriving the component Lagrangian, the Hamiltonian and quantizing the latter, we obtain a certain supersymmetric 
quantum mechanical spectral problem, which is isomorphic to the Dirac spectral problem. It involves the gauge potentials
$A_{w,\bar w} = (-i\partial G, i\bar \partial G)$ and the magnetic field 
 $F_{w \bar w} = -F_{\bar w  w} = 2i \partial \bar \partial G$. 
  Let us choose 
 \be
\label{FWW}
G  \ =\ - \frac q2 \ln(1 + \bar W W) \ .
 \ee
Then the magnetic field area density
 $(1+\bar w w)^2 F_{w \bar w} $ is constant on $S^2$ and $q$ is the value of the flux (\ref{IndS2}).
 The quantum supercharges 
(isomorphic to $/\!\!\!\!{\cal D}(1 \pm \sigma^3)/2$) 
are 
 \be
\label{QQbar}
Q &=& i (1 + \bar w w) \psi \left[ \partial - \frac {\bar w (1-q)}{2(1 + \bar w w)} \right] \nonumber \\
\bar Q &=& i (1 + \bar w w) \bar\psi \left[ \bar \partial - \frac { w (1+q)}{2(1 + \bar w w)} \right] \, .
  \ee
They act upon the two-component wave functions 
 \be
 \label{Psicov2comp}
\Psi^{\rm cov} \ =\ \Psi_{F=0} (\bar w, w) + \psi \Psi_{F=1} (\bar w, w)
 \ee
 normalized with the covariant measure 
 \be
\label{measure}
\mu \, d\bar w dw\ =\   \sqrt{g} \, d\bar w dw \  =\  \frac {d\bar w dw}{(1 + \bar w w)^2}\, ,
 \ee
($ F = \psi \bar \psi = \psi \frac {\partial}{\partial \psi}$  is the operator of the fermion charge  
commuting with the Hamiltonian.) 
 The operators $Q, \bar Q$ are (naively) Hermitially conjugate one to  another with the measure (\ref{measure}), 
$\bar Q = \mu^{-1} Q^\dagger \mu$. The supercharges (\ref{QQbar}) are nilpotent and $\{\bar Q, Q\}$ is the Hamiltonian.
In the sector $F=0$, the latter reads
\be
\label{HF0}
H_{F=0} \ =\ -(1 + \bar w w)^2 \partial \bar \partial + \kappa (1 + \bar w w) (\bar w \bar \partial - w \partial)
+ \kappa^2 \bar w w + \kappa \ ,
 \ee
where 
  \be
 \label{kappa}  
\kappa = \frac {1-q} 2 \ .
  \ee
It commutes with the angular momentum operator $ L =  \bar w \bar \partial - w \partial$. Note that the 
Hamiltonian (\ref{HF0}) coincides up to a constant shift with the Hamiltonian for the scalar charged
particle in the field of a monopole with the magnetic charge $q' = q-1$.  

The expression for $H_{F=1}$ is similar, one has only to interchange $w$ and $\bar w$ and inverse the sign of $q$.  

Acting with the 
hamiltonian (\ref{HF0}) on the wave function $\Psi = \bar w^m F(\bar w w)$ in the sector with a given angular momentum $m$,
we obtain the equation 
 \be
\label{eqlu}
-u(1+u)^2 F''(u) - (m+1) (1+u)^2 F'(u) + \kappa^2 u F(u) + \kappa m (1+u) F(u) \nonumber \\
 = (\lambda - \kappa) F(u) \ ,
 \ee
where $u =\bar w w$ and $\lambda$ is the spectral parameter. Introducing the variable
 \be
\label{zu}
z \ =\ \frac {1-u}{1+u}
 \ee
(it is nothing but $\cos \theta$, $\theta$ being the polar angle on $S^2$), we derive
 \be
\label{eqlz}
(z^2 -1) F''(z) + 2 (z+m) F'(z) + \frac {2\kappa(\kappa + m)}{1+z} F \ =\ [\lambda - \kappa(1-\kappa) ] F(z) \ .
 \ee
There are two families of formal solutions:
 \be
\label{solFz}
F_n^m(z) &=& (1+z)^{-\kappa} P^{m,-m-2\kappa}_n(z) \ \ \ \ \ \ \  [\lambda_n = n(n+1 - 2\kappa)]\, ,  \nonumber  \\
\tilde F_n^m(z) &=& (1+z)^{\kappa +m} P^{m,m+2\kappa}_n(z) \  \ \ \ [\tilde \lambda_n = (m +n +2\kappa)(m+n+1)] \ ,
 \ee
where $P^{\alpha,\beta}_n(z)$ are the Jacobi polynomials,
 \be
\label{Jacobi}
P^{\alpha,\beta}_n(z) \ =\ \frac 1{2^n} \sum_{k=0}^n \left( \begin{array}{c} n+\alpha \\ k \end{array} \right)
 \left( \begin{array}{c} n+\beta \\n- k \end{array} \right) (1 + z)^k (z-1)^{n-k} \ .
 \ee
For $\alpha > -1,\ \beta > -1$, the Jacobi polynomials are mutually orthogonal 
on the interval $z \in (-1,1)$ with the weight
$\mu = (1-z)^\alpha (1+z)^\beta$.

An important observation is that, for integer $q$, one has not two families of solutions, but actually only {\it one} such family.
Indeed, for integer $\alpha, \beta$, the Jacobi polynomials satisfy  the following interesting relations  \cite{WuYang}
  \be
\label{relJacobi}
P^{-\alpha, \beta}_{n+\alpha} = 2^{-\alpha} (z-1)^\alpha \frac {n! 
(n+\alpha + \beta)!}{(n+\alpha)! (n+\beta)!} P_n^{\alpha, \beta} \ ,
\nonumber \\  
P^{\alpha, -\beta}_{n+\beta} = 2^{-\beta} (z+1)^\beta \frac {n! 
(n+\alpha + \beta)!}{(n+\alpha)! (n+\beta)!} P_n^{\alpha, \beta} \ .
 \ee
Using the second relation, it is easy to see that, for integer $q$,
 the functions $F_n^m(z)$ and $\tilde F_n^m(z)$ coincide up to a shift of $n$ and  an irrelevant factor. 
Picking up only the square integrable functions, we 
obtain the so called monopole harmonics \cite{WuYang,KimLee},
  \be
  \label{monharm}
  \Psi^{(q)\, F=0}_{mn} = e^{i m\phi} (1- z)^{|m|/2} (1+z)^{|m+2\kappa|/2}  P_n^{|m|, |m+2\kappa|}(z) \, ,   \nonumber \\
   n = 0,1,\ldots \,  (q > 0), \ \ \ \ \ \ \ \ \ n = 1-q, \ldots \, ( q \leq 0) \ , \nonumber \\
  m = -n, \ldots, n+q -1 
   \ee 
with $e^{i\phi} = \sqrt{\bar w/w}$ (we do not bother about the normalization coefficients).  The spectrum is
    \be
 \label{lamF0}
 \lambda_{mn}^{(q)\, F=0} \ =\ ( | m + \kappa | + n + \kappa ) (|m + \kappa| + n +1 -\kappa)\ .
   \ee
 The eigenfunctions and the eigenvalues in the sector $F=1$ are given by the same expressions with the 
change $q \to -q, \ m \to -m$ 
(the eigenfunctions involve, of course,  the extra fermion factor $\psi$).

 If choosing $q=1$ in the sector $F=0$ (or $q = -1$ in the sector $F=1$), the expressions are simplified. In this case,
 $\kappa=0$ and the Hamitonian (\ref{HF0}) is reduced to the ordinary
Laplacian\footnote{A mathematician would remark that, in this case, the twisted Dirac complex is equivalent to the untwisted Dolbeault
($q=1$) or  
anti-Dolbeault ($q=-1$) complex. See e.g. \cite{IvSm} for more detailed discussion.}.
 The eigenfuctions represent then the Gegenbauer polynomials. They are all regular on $S^2$. The spectrum is $\lambda = l(l+1)$
(with $l = n+|m|$) as it should be.

 For other {\it integer} $q$, the situations is somewhat more complicated. There are  functions with $m+2\kappa = 0$ and
$m \neq 0$ which behave as $\sim e^{im\phi}$ in the vicinicy of the north pole and are nothing but the sections of the fiber
bundle associated with the gauge field. All monopole harmonics (\ref{monharm}) are square integrable on $S^2$. 

 One can make now two important observations:
\begin{itemize}
\item For integer $q$, one can be convinced that the action of the supercharge $Q$ on  a square integrable
 eigenfunction (\ref{monharm})  
 is also square integrable. The same concerns the action of $\bar Q$ on the square integrable eigenfunctions of $H_{F=1}$. 
In other words, the Hilbert space spanned
by the functions (\ref{monharm}) in the two fermionic sectors lies in the domain of $Q$ and $\bar Q$, and hence in the domain
of \ \  $/\!\!\!\!{\cal D}$. 
As a result, the full spectrum of the hamiltonian $H = \ /\!\!\!\!{\cal D}^2$ is supersymmetric --- 
all excited states are doubly degenerate.
 \item The Witten index $n^{(0)}_{F=0} - n^{(0)}_{F=1}$ of this system (alias, the Atiyah-Singer index of\ $/\!\!\!\!{\cal D}$) 
is equal to $q$.
  \end{itemize}

\section{Fractional charge}

Let us see now what happens if $q$ is not integer. The first immediate observation is that the functions $\tilde F^m_n$
are not expressed via  $F^m_n$ anymore, and we have {\it a priori} two different families of solutions.

As an example, consider the case $q = 1/2$. The solutions are 
 \be
\label{2famq12}
 \Psi_{mn}^{(1/2) F=0} &=& e^{im\phi} (1-z)^{m/2} (1+z)^{-1/4-m/2} \, P_n^{m, -m - 1/2}(z), \nonumber \\
 \tilde \Psi_{mn}^{(1/2) F=0}  &=& e^{im\phi} (1-z)^{m/2} (1+z)^{1/4+m/2} \, P_n^{m, m + 1/2}(z) \, .
 \ee
We are interested, however, only in normalizable solutions. Thus, when $m > 0$, only the second set of solutions 
$  \tilde \Psi_{mn}$ is admissible, while the normalization integral $\int   |\Psi_{mn}|^2 dz $ diverges ar $z=-1$.
 When $m <0$, it may seem at first that there are no normalizable solutions whatsoever
due to the divergence at $z=1$. Well, this divergence is there for $n < |m|$. But for larger $n$, one can use
 the first relation in Eq.(\ref{relJacobi}) and express 
 \be
\label{tipo19}
P_n^{-|m|, \pm |m| \mp 1/2}(z)  \propto   (1-z)^{|m|}P_{n - |m|}^{|m|, \pm |m| \mp 1/2}(z) \, .
  \ee
 Then the divergence at $z=1$ (that corresponds to $w = \bar w = 0$)  disappears
and we have only to take care about the divergence at $z=-1$ (or $w = \bar w = \infty$). For $m \leq -2$, it is always there
for the  family $  \tilde \Psi_{mn}$ and only the family $\Psi_{mn}$ is admissible. All
 these normalized solutions are regular on $S^2$.

The values $m = 0, -1$ are special, however.
In both cases, there are {\it two} normalizable families:
 \be
\label{2famm0}
\Psi_{0n}, \tilde \Psi_{0n} \  =  \ 
(1+z)^{\mp 1/4} P_n^{0, \mp 1/2}(z) \, , \nonumber \\
 \Psi_{-1,n}, \tilde \Psi_{-1,n} \ = \ e^{-i\phi} (1+z)^{\pm 1/4} (1-z)^{1/2} P_{n-1}^{1, \pm 1/2}(z) 
 \ee
(we used the property (\ref{tipo19}) in the second line). 
Half of these states are singular on $S^2$, while another half are not.

We can see now that the Hamiltonian (\ref{HF0}) is not Hermitian in the Hilbert space including all the 
functions in (\ref{2famm0}). Indeed, consider the sector $m=0$ and restrict ourselves with the case $n=0$. There are two 
normalizable states,
 \be
\label{2states}
|1 \rangle &=& (1+z)^{1/4} \ \ \ \ {\rm with} \ \lambda = \frac 12 \, , \nonumber \\  
|2 \rangle &=& (1+z)^{-1/4} \ \ \ \ {\rm with} \ \lambda = 0 \, .  
  \ee
The states $|1,2 \rangle$ have different eigenvalues, but are {\it not} orthogonal to each other, 
$\langle 1 | 2 \rangle \neq 0$. 
This means that the hermiticity of (\ref{HF0}) is lost. One can also see that, if comparing the matrix elements
$\langle 1 | H | 2 \rangle$ and  $\langle 2 | H | 1 \rangle$, they differ by an integral of a total derivative
\footnote{
A similar phenomenon can be observed in other settings. Consider for example the covariant Laplacian on $S^3$. Its explicit
expression in stereographic coordinates  is
 \be
\label{LaplS3}
H = - \triangle_{S^3} \  =\ -  f^3 \partial_i \frac 1f \partial_i 
 \ee
with $f = 1 + r^2/4$. (The metric is then $ds^2 = d\vec{x}^2/f^2$, the radius of such 3-sphere being $R=1$.) 
This operator has some set of regular on $S^3$ eigenfunctions, in particular the eigenfunction $\Psi_1 = 1$ with 
$\lambda =0$. One can try to consider, however, the operator (\ref{LaplS3}) on $S^3 \backslash $\{{\sl pole}\} 
and include into the consideration also
singular at infinity functions with the only requirement for them to be square integrable.
There is one such singular square integrable  eigenfunction of $H$:  $\Psi_2 =\sqrt{f}$ with the eigenvalue $\lambda = -3/4$. 
(Recall that the normalization integral includes the measure $\mu = \sqrt{g} = 1/f^3$.)  As the eigenvalues
are different and $\langle 1|2 \rangle
\neq 0$,  hermiticity in the extended Hilbert space that includes $\Psi_2$ is lost. One can also compare 
$\langle 1 | H | 2 \rangle$ and  $\langle 2 | H | 1 \rangle$ and find out that 
 \be
\langle 1 | H | 2 \rangle -  \langle 2 | H | 1 \rangle \ 
\sim \int d^3x \, \partial_i \left[ \frac {x_i} {r^3} \right] \neq 0\, .
 \ee
 }

 \be
\langle 1 | H | 2 \rangle -  \langle 2 | H | 1 \rangle \ \sim \int \partial \left[ \frac w {1 + \bar w w} \right] dw d\bar w
\ \sim  \int d^2x \, \partial_i \left[ \frac {x_i} {r^2} \right] \neq 0\, .
 \ee

\begin{figure}
\label{ecelqF0}
 \begin{center}
        \epsfxsize=270pt
        \epsfysize=0pt
        \vspace{-5mm}
        \parbox{\epsfxsize}{\epsffile{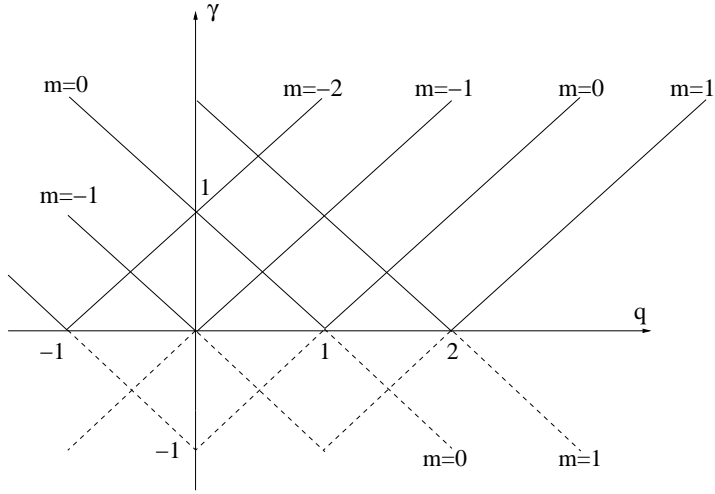}}
        \vspace{5mm}
    \end{center}
\caption{The eigenstates of (\ref{HF0}) for different $q$. Solid lines --- regular functions. Dashed lines --- singular but square integrable functions.}
\end{figure}

\begin{figure}
\label{ecelqF1}
 \begin{center}
        \epsfxsize=270pt
        \epsfysize=0pt
        \vspace{-5mm}
        \parbox{\epsfxsize}{\epsffile{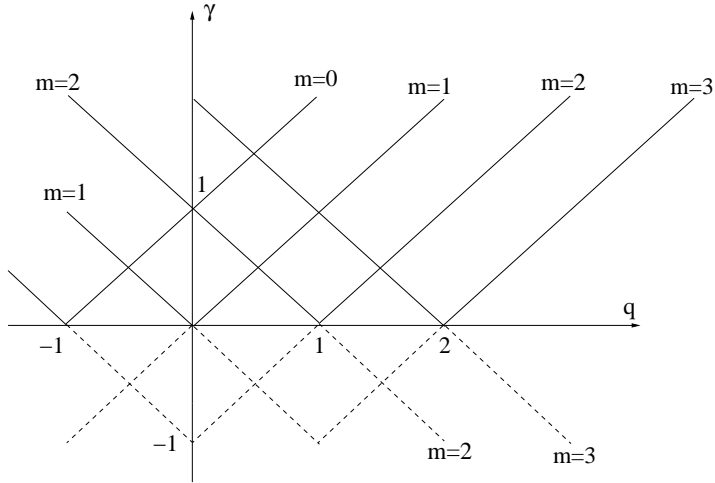}}
        \vspace{5mm}
    \end{center}
\caption{The same in the sector $F=1$.}
\end{figure}

This analysis can be generalized for  other values of $q$. 
In Fig. 1, we plotted the power 
$\gamma$ in the asymptotic behavior of the wave functions at infinity 
,  $\Psi_{mn}^{(q)} \sim |w|^{-\gamma}$.
Note that it is {\it not} the energy that is plotted, such that a crossing of the lines in Fig. 1
does not generically mean a degeneracy of the levels. However, at $\gamma = 0$, each crossing  
involves a {\it single} tower of states
(there is also the quantum number $n$ that 
marks the levels of these towers)
 rather than two different towers.

Positive $\gamma$ correspond to nonsingular at the north pole functions. The functions
 with $-1 < \gamma < 0$ are singular but normalizable, while the functions with
$\gamma \leq -1$ are not normalizable. One can observe that two 
extra singular normalizable  towers of states  in the sectors $m=0, -1$ 
exist not only for $q=1/2$, but for any $ q \in (0,1)$. 
For other noninteger $q$, the two extra towers are present in a different pair of sectors. 
For example, if $q= 2.7$, these are the
sectors $m = 1,2$.      

 One can also perform a similar analysis for the Hamiltonian in the sector $F = 1$.  
The structure of the levels shown in Fig. 2 is the same as in the sector $F = 0$ up to the shift $m \to m+2$. 
This means that, e.g.
for $q = 1/2$, we have two extra troublesome normalizable singular towers of states in the sectors $m = 1,2$, etc.   
 
Let us try now to redefine a problem and {\it exclude} the singular functions from consideration.
Note that the functions (\ref{2famm0}) either grow or vanish at infinity, the section-of-a-bundle behavior $\sim e^{im\phi}$
characteristic for monopole harmonics with integer $q$ is not possible here. Thus, our Hilbert space includes only regular on $S^2$
functions.  
The Hamiltonian becomes in this case
Hermitian in the both fermionic sectors.
\footnote{Thus, the title of this paper implying that noninteger flux does {\it not} work is not quite exact. To some
extent, it {\it works}. This concerns especially the nonsupersymmetric problem of a scalar charged particle in the 
magnetic monopole field. It is benign even if the monopole charge is not integer.} 
 The trouble strikes back, however, when we try to act on the states in this reduced 
Hilbert space by the supercharges. It particular, let us act by the supercharge $Q$ on the states with $F=0$.
Let first $q=1/2$ and pick up the lowest ($n=0$)  nonsingular states
   \be
\label{F0nesing}
 \Psi_{00}^{F=0} \ =\ \frac 1{(1 + \bar w w)^{1/4}} \, , \ \ \ \ \ \ \ \ \ \ \ 
 \Psi_{-1,0}^{F=0} \ =\ \frac w{(1 + \bar w w)^{3/4}} 
\ee
We derive
  \be
\label{QPSI}
 Q  \Psi_{00}^{F=0} \ =\ \Psi_{10}^{F=1} \ =\  \psi \frac {\bar w}{(1 + \bar w w)^{1/4}} \, , \ \ \ \ \ \ \ \ \ \ \ 
 Q \Psi_{-1,0}^{F=0} \  =\ \Psi_{00}^{F=1} \ =\ \psi \frac 1{(1 + \bar w w)^{3/4}} 
\ee
The second function in Eq.(\ref{QPSI}) 
is regular at $w = \infty$, whereas the first 
one is normalizable but singular !

Generically, one can observe that
 the lines with positive slope in Fig. 1 are shifted  
upwards upon the action of $Q$ (such that the parameter $\gamma$ is increased, increasing the convergence 
of the normalization integral.) 
On the other hand, the lines with  negative slope are  shifted downwards. This means that  some nonsingular
 functions become singular  
 (and roughly a half of singular nonrenormalizable functions become normalizable). And that means that the
 supersymmetry of the spectrum is lost. Some eigenfunctions of the Hamiltonian do not have superpartners. 

The fact that the action of the supercharge can make a singular function out of a nonsingular one
 is rather natural, bearing in mind the generic structure of (\ref{QQbar}) at large $w$, $Q \sim \bar w w (\partial + 1/w) 
\sim \bar w$. A special explanation may rather be required why this does {\it not} happen for integer $q$. 
Well, the reason is
that  the problematic  function $\Psi$ with positive $\gamma$ and negative slope giving singular  $ Q\Psi$ fuses 
at integer $q$ 
with a positive slope function for which  $ Q\Psi$ {\it is}  not singular.

A similar phenomenon (the loss of apparent supersymmetry due to the fact that the superpartners of some states lie outside 
the Hilbert space and thus do not belong to the spectrum of the Hamiltonian) 
is known to show up for some other systems. The simplest example \cite{SSV}
 is probably Witten's supersymmetric Hamiltonian
 \be
\label{WitHam}
 H = \frac {p^2}2 + \frac {[W'(x)]^2}2 + \frac {W''(x)}2 [\bar \psi, \psi] 
 \ee 
with the superpotential $W' = -\omega x + 1/x$.
In the bosonic sector $F=0$, the Hamiltonian is 
 \be
\label{HB}
 H_B = \frac {p^2}2 + \frac {\omega^2 x^2 }2 - \frac {3\omega}2 
 \ee
The ground state $\Psi_0(x) \propto \exp\{- \omega^2 x^2/2 \} $ has the negative energy $E = -\omega$, which
obvously is not consistent with supersymmetry. The reason for the trouble is that the function $\Psi_0$ does not belong
to the domain of the supercharge $Q \propto p + iW'$, $ Q\Psi $ being not square integrable. To make the
spectrum supersymmetric, we should in this case restrict the Hilbert space and consider only the functions that
vanish at the origin, $\Psi(0) = 0$. With this restriction, the ground state $\tilde \Psi_0(x) \ = \ x \Psi_0(x)$
has zero energy.

On the other hand, for a magnetic field with noninteger flux, 
there is no way to make the Pauli Hamiltonian  supersymmetric. We have just shown that the Hilbert space of 
regular functions on $S^2$ does not constitute the domain of $Q$, but this is true also for any restricted
or enhanced Hilbert space. If we start, for example, from the space of square integrable functions, we can  easily
see that the action of the
supercharge on the singular square integrable functions marked with dashed lines in Figs. 1,2 would produce 
the  functions for which the normalization integral diverges.

The last comment we want to make is the following. We have seen that the Hamiltonian (\ref{HF0}) is not Hermitian 
in the Hilbert space of square integrable functions. However, the spectrum of this Hamiltonian is real and it is known
that such Hamiltonians belong to the class of quasi-Hermitian (or crypto-Hermitian \cite{crypto}) Hamiltonians 
that can be rendered real, if redefining the inner product in a special way \cite{quasi}. It is also possible
to do for the Hamiltonian (\ref{HF0})) (and also for the more simple Hamiltonian (\ref{LaplS3})), if defining the inner
product in a usual way $ \langle 1| 2 \rangle  =\int \Psi_1^* \Psi_2 \, \mu \, d^2x$ for  nonsingular 
on $S^2$ functions, but postulating 
that extra singular normalizable states are orthogonal to nonsingular ones. No doubt, such an inner product is very unnatural,
but it can in principle be chosen and then the Hamiltonians (\ref{HF0}) and (\ref{LaplS3}) would be Hermitian.

\section{Acknowledgements}
I am indebted to E. Ivanov and A. Wipf for useful discussions.

\end{document}